# Theoretical Framework and Simulation Results for Implementing Weighted Multiple Sampling in Scientific CCDs

**Cristobal Alessandri**[1], **Dani Guzman**[1], **Angel Abusleme**[1], **Diego Avila**[1], **Enrique Alvarez**[1], **Hernan Campillo**[1], **Alexandra Gallyas**[1], **Christian Oberli**[1], **Marcelo Guarini**[1]

*[1] Dept. of Electrical Engineering, Pontificia Universidad Católica de Chile, Santiago, Chile*

**ABSTRACT**

*The Digital Correlated Double Sampling (DCDS) is a technique based on multiple analog-to-digital conversions of every pixel when reading a CCD out. This technique allows to remove analog integrators, simplifying the readout electronics circuitry. In this work, a theoretical framework that computes the optimal weighted coefficients of the pixels' samples, which minimize the readout noise measured at the CCD output is presented. By using a noise model for the CCD output amplifier where white and flicker noise are treated separately, the mathematical tool presented allows for the computation of the optimal samples' coefficients in a deterministic fashion. By modifying the noise profile, our simulation results get in agreement and thus explain results that were in mutual disagreement up until now.*

## 1. INTRODUCTION

Charge-Coupled Devices (CCD) have been one of the greater advances in astronomy since they were popularized in the 90's [1]. A CCD generates electrons from absorbed and detected incoming photons, which have to be converted to a measurable voltage at the CCD's output amplifier. Current CCDs have very low noise amplifiers, and their output signals are processed by carefully designed, low-noise readout electronics. Since its inception in astronomy, the lowest noise in CCDs have been obtained by filtering each pixel signal using an analog integrator to reduce white noise, and a dual readout scheme to cancel reset noise.

The resolution of the Analog-to-Digital Converter (ADC) for astronomical images is typically 16 bits, to allow proper sampling of the full dynamic range of the pixel well while still detecting faint intensity variations. Up until recent years, 16 bits ADC could not run much faster than a few tens of Megahertz, limiting the option of multiple sampling of every pixel while still keeping the whole readout period to a short time. Nevertheless, it is now possible to run very fast 16 bits ADC[1] which will allow a digitized CCD output, where more complex filtering than a simple integrator could be implemented in an attempt to reduce the noise below

---

[1] *Analog Devices' fastest 16 bits ADC is AD9467 and achieves up to 250 MSPS.*

the floor that an analog integrator can reach.

It is known for decades [2] that a simple integrator is the best filter when the noise is dominated by a white component. Under the scenario of digital multiple sampling, this optimal filter can be expressed as a simple average of the samples. In our scheme of coefficients multiplying each sample, all coefficients take an identical value.

There have been attempts to reduce the noise applying different coefficients to the multiple sampling scheme. Gach [3] experimentally and somewhat intuitively found better weights for a particular CCD, reducing the readout noise from 5 to 1.7 electrons. Clapp presented experimental results with different weighting coefficients, finding out that constant (or flat) weights were optimal for his CCD [4]. Clapp even tried to apply Gach's coefficient profiles, but could not find anything better than flat weights. In this sense, both results have been considered in disagreement.

In this context, both studies were purely experimental, but no analytical tools were used to justify the weights profile selected. In this paper, an analytical method to determine the optimal weighting is proposed and validated with simulation results.

## 2. METHOD

Fig. 2-1 shows a simplified model to compute the output-referred noise contribution of a CCD readout system using DCDS. The CCD output amplifier transfer function is given by $H(s)$, and it may also consider the external electronics. Any white or colored white noise source can be modeled in the time domain as a sequence of noise pulses with fixed amplitude and shape but random sign [5]. The core pulses $y(t)$ that represent the noise process $\overline{n^2}(f)$ can be calculated as shown in [5]. The effect of each individual noise pulse at the sampler input can be calculated by the convolution of $y(t)$ and the CCD's impulse response as shown in (2-1). Fig. 2-2 shows the pulse shape of $\hat{y}(t)$ at the sampler input for white and flicker noise sources, assuming a single pole transfer function $H(s)$ with time constant $\tau$.

$$\hat{y}(t) = y(t) * \mathcal{L}^{-1}\{H(s)\}(t) \qquad (2\text{-}1)$$

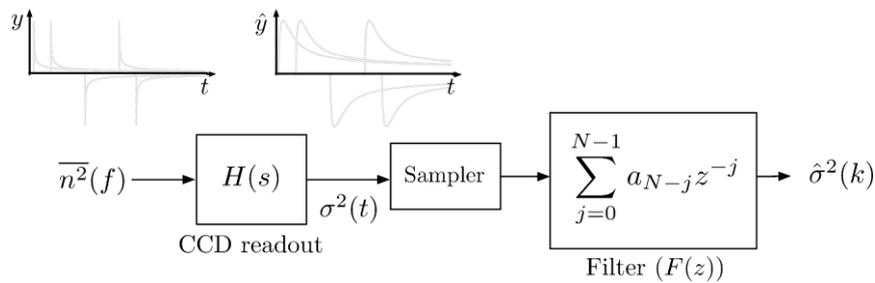

*Figure 2-1  Model of a CCD readout system encompassing a transfer function $H(s)$ which can include the CCD output amplifier as well as external electronics; an ideal sampler (i.e. an ADC with infinite resolution) and a discrete filter.*

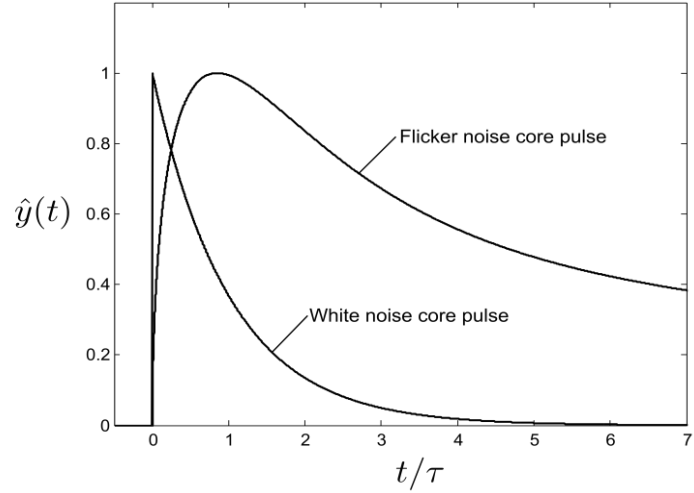

*Figure 2-2 White and flicker normalized pulse shapes at the sampler input [5].*

The total RMS noise at the sampler input is a function of time (i.e. $\sigma^2(t)$), and can be calculated as shown in [5]. However, the total output RMS noise $\hat{\sigma}^2(k)$ cannot be calculated directly with a discrete convolution because different samples hold partial correlation from pulses originated at earlier samples.

The noise analysis method for particle physics experiments presented in [6] allows to obtain a closed-form expression for the total integrated noise at the filter output as follows:

Calculate the noise contribution of pulses generated during an interval $P_i$ and measured at an arbitrary sample $k$. This contribution can be expressed as (2-2) and is graphically shown in Fig. 2-3 [6].

$$\sigma_i^2(k) = \sigma^2\big((k-i+1)T_s\big) - \sigma^2\big((k-i)T_s\big) \quad k \geq 1 \quad (2\text{-}2)$$

The total integrated noise at the sampler input for a sample $k$ can be written as the sum of individual noise contributions originated at each interval $P_i$, as shown in (2-3) and Fig. 2-4 [6].

$$\sigma^2(k) = \sum_{i=1}^{N} \sigma_i^2(k) \quad (2\text{-}3)$$

Since $\sigma^2(k)$ is composed by noise contributions originated at different time intervals $P_i$, the output noise cannot be computed directly by a discrete convolution. On the contrary, all evaluations of $\sigma_i(k)$ are originated from the same pulses (for a fixed $i$), so $\sigma^2(k)$ can be split into $N$ independent discrete-time signals and referred to the output as shown in (2-4) [6].

$$\hat{\sigma}^2(k) = \sum_{i=1}^{N} \left( \sqrt{\sigma_i^2(k)} * Z^{-1}\{F(z)\}(k) \right)^2 \quad (2\text{-}4)$$

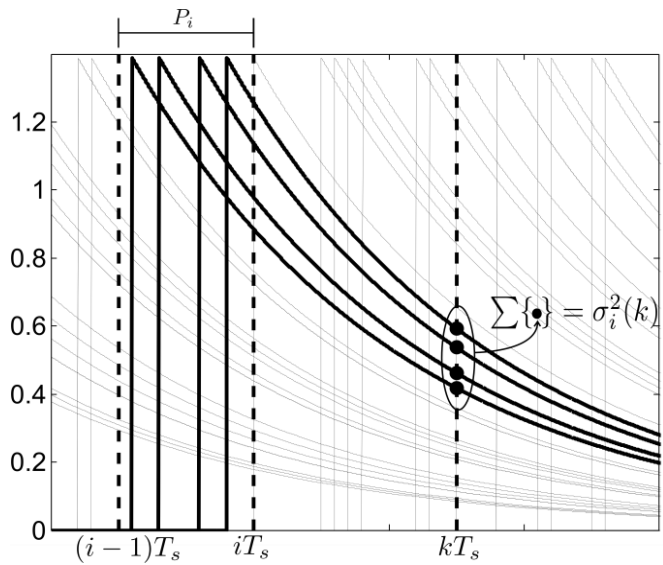

*Figure 2-3 Noise contribution of pulses generated within $P_i$ and measured at an arbitrary sample $k$ for an arbitrary filtered noise core function $\hat{y}(t)$. The total contribution is the sum of black dots in quadrature [6].*

This method was generalized to consider noise sources originated from an arbitrary time. Then it was adapted and applied to a CCD DCDS readout system. Calculating the noise contribution of Gaussian, flicker and reset noise, optimal weights can be obtained to reduce the readout noise using standard optimization techniques.

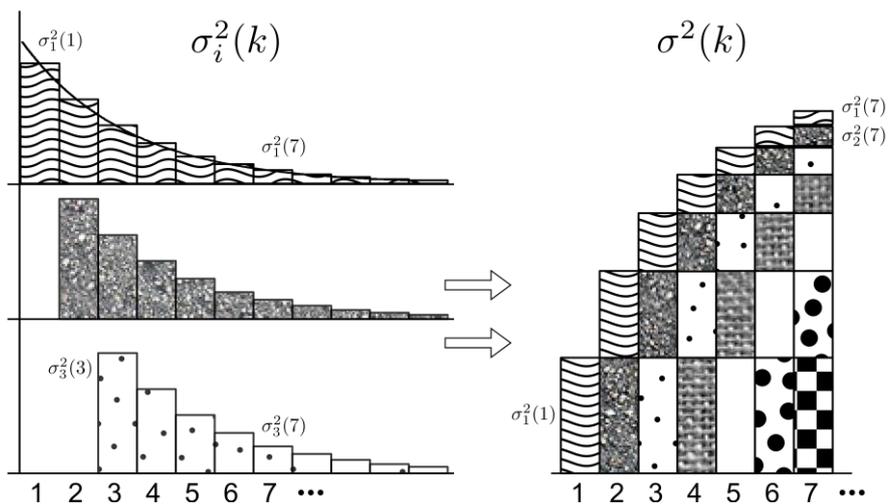

*Figure 2-4 Evolution of the total integrated noise at the filter input, where the noise of each sample was split according to (2-3) [6].*

## 3. SIMULATION RESULTS

For validation purposes the CCD readout system was simulated using MATLAB. White and flicker noise sources with known power spectral density were generated in the time domain. These noise sources were filtered by the transfer function of a CCD $H(s)$.

The method was simulated with parameters extracted from e2v's CCD231-84[2] datasheet and Clapp's experimental results [4]. On each case, system parameters such as noise floor, corner frequency and bandwidth were adjusted to match the RMS noise vs pixel frequency plot for an averaging filter at a sampling frequency of 2 MHz. The solid lines in Figs. 3.1 and 3.3 show the extracted noise data, whereas the dotted lines show the simulation results using the estimated parameters.

Once the parameters are known, the method can be applied calculating the optimal filter coefficients once for each CCD and pixel time. The simulation was executed again modifying only the filter weights.

The circles in Fig. 3.1 show the resulting noise from our simulations using the set of weights plotted in Fig. 3.2, whereas the circles in Fig 3.3 show the results using the weights plotted in Fig. 3.4.

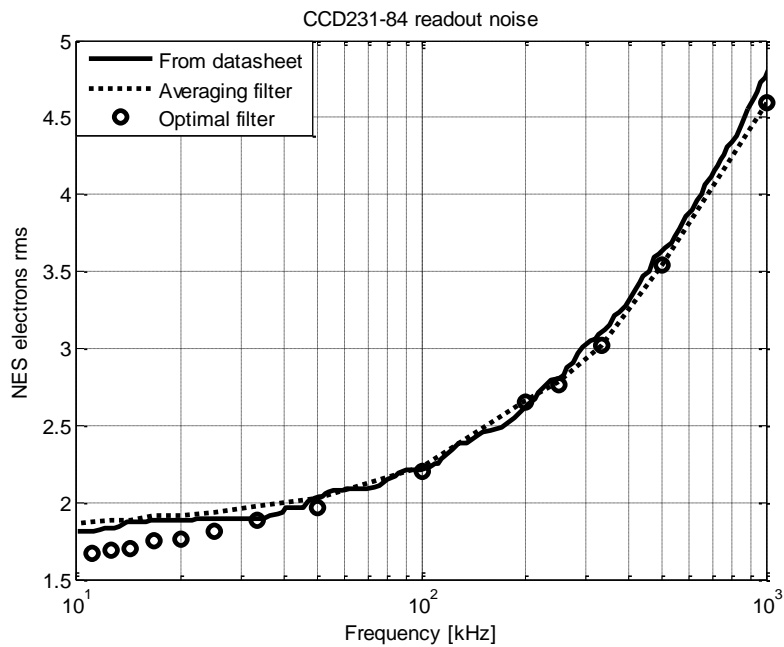

*Figure 3-1 CCD231 readout noise comparison: from datasheet, simulated using an averaging filter and with optimal weights.*

---

[2] *Available at www.e2v.com.*

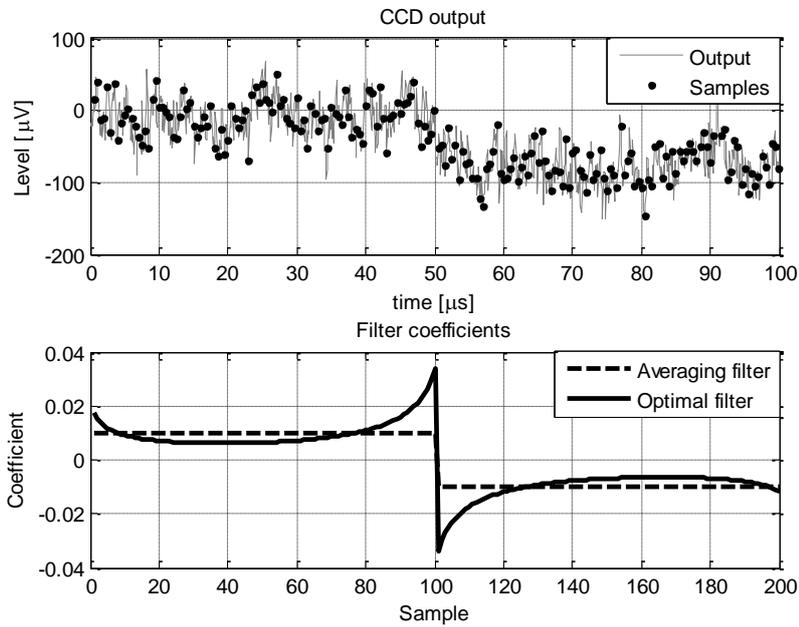

*Figure 3-2 Simulated CCD output video waveforms (top) and Optimal weights for CCD231 at 100µs pixel time (bottom).*

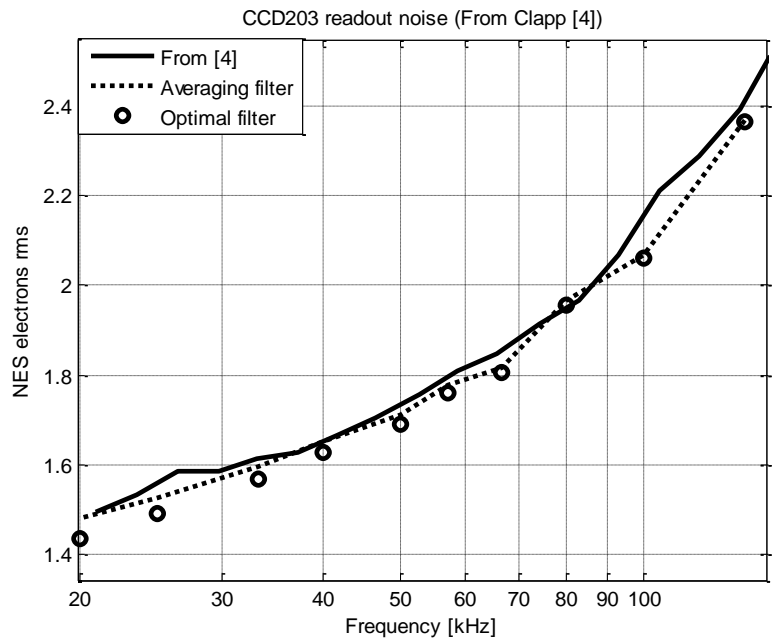

*Figure 3-3 CCD203 readout noise comparison: from [4], simulated using an averaging filter and with optimal weights.*

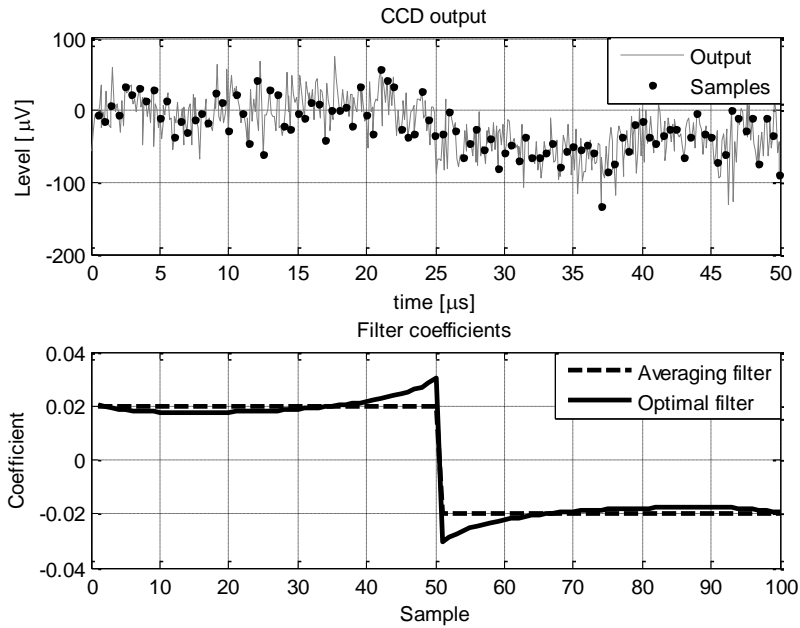

*Figure 3-4 Simulated CCD output video waveforms (top) and Optimal weights for CCD203 at 100µs pixel time (bottom).*

For CCDs with low corner frequency, the optimal coefficients are the same as those in an averaging filter. For higher flicker noise contribution, samples around the beginning and the end of each integration period are assigned larger weights and the optimal coefficients are not perfectly symmetric.

## 4. CONCLUSIONS

This paper presents a method that allows for the determination of the optimal coefficients for a given CCD setup knowing its noise floor, corner frequency and bandwidth. The RMS noise obtained with the optimal filter is always lower or equal to that resulting from using an averaging filter.

The shape obtained for low corner frequency (i.e. dominated by white noise) agrees with Hegyi's results. For high corner frequency (i.e. dominated by flicker noise), the optimal filter shape is very similar to the coefficients found by Gach, although larger weight is assigned to samples at the beginning and at the end of each pixel. This difference can be understood when studying the original framework developed by Avila to model the noise in the time domain, but it is beyond the scope of this paper to describe it.

The method presented in this work shows a better performance for both CCDs, achieving up to a 12% noise reduction. The improvement is more significant for slow pixel rates, where a flat-weighted filter fails to reduce the flicker noise contribution.

## 5. ACKNOWLEDGEMENTS

The authors appreciate support from FONDECYT-CONICYT program under grant 1130334.